# Statistical approach of random packing of polydisperse hyperspheres: with application to binomial and lognormal size distributions

H.J.H. Brouwers
*Department of the Built Environment, Eindhoven University of Technology*
*P.O. Box 513, 5600 MB Eindhoven, The Netherlands*
*Contact address:jos.brouwers@tue.nl*

This paper addresses the packing fraction of polydisperse similar hyperspheres, with application to discrete binomial and continuous lognormal particle size distribution. First, a closed form expression is derived of a mix consisting of multiple discrete particles in D-dimensional Euclidean space $\mathbb{R}^D$ with small size differences. A recently developed analytical model for binary packings with small size ratio u is used, that has its foundations in the excluded volume model by Onsager, here applying it to the statistically probable particle pairs in the multi-sized mix. The new model is subsequently applied to discretely sized particles with a binomial number distribution and a constant size ratio u between each subsequent particle size. By taking the limit of u → 1 and an infinite number of particles sizes, the binomial particle distribution is transformed into a continuous lognormal particle size distribution. The resulting expression for the packing fraction can be solved in closed form, resulting in an original expression for the packing fraction of mixtures of hyperparticles with a lognormal distribution. This explicit expression is governed by the standard deviation, the Euclidean dimension D, and the monosized packing fraction of the considered particle shape, the latter depending on mode of packing and particle shape only. For a number of particle shapes and their packing modes (from looser to denser) the scaled expression is thoroughly compared with empirical and computational data reported in the literature. They concern packed particles with small standard deviations and in Euclidean dimensions D = 2 and 3, and good agreement is found. For large standard deviations the scaled packing fraction approaches a limiting value, which appears not to depend on the Euclidean dimension. Based on the insight, a closed form expression is put forward for the packing fraction that matches the small standard deviation solution, and the large standard deviation asymptote, so covering arbitrary standard deviations, Euclidean dimensions, particle types, and their mode of packing.

## 1. INTRODUCTION

The packing of polydisperse particles is a fundamental problem in particulate materials and occurs in applications ranging from powders and ceramics to soils, pharmaceuticals, concrete, and composite materials [1]. The packing fraction of a particulate assembly determines its void fraction and strongly influences properties such as permeability, mechanical behavior, transport, shrinkage, and the amount of binder or matrix required in composite systems. The optimal packing of hard spheres in higher dimensions is of interest in error-correcting codes in communication theory [2]. Furthermore, attention has been paid to revealing the packing geometries and the route to understanding liquids, colloids, glasses, metals and metalloids [3].

In contrast to monodisperse packings, in which all particles have the same size, polydisperse systems contain particles of different sizes that can occupy different parts of the available space. The resulting packing fraction is therefore governed by the particle-size distribution as well as the mode of packing (the densification). When equally shaped particles with different sizes are randomly packed, i.e. generating a polydisperse packing of similar particles, the packing fraction increases compared to the monosized packing of the congruent (or identical) particles. Furthermore, polydispersity also inhibits crystallization and enables vitrification [3], but the impact of polydispersity on glass transition has yet to be fully elucidated. So there is practical as well as fundamental interest in understanding the relationship between the particle shape and particle-size distribution on the one hand, and its effect on packing fraction and glass transition on the other.

The fundamental role of particle size distribution on packing fraction was recognized in early studies of granular mixtures [4-11]. Later, the effect of polydispersity on packing and glass transition has for instance been studied experimentally [12-16], numerically [2, 17-25], and theoretically (replica theory and mode coupling theory) [3, 26-30].

A theoretical treatment of hard particles, using the second coefficient from the virial expansion of the pressure, provides a natural alternative. For hard particles, this second virial coefficient is equal to half the orientationally averaged excluded volume between two particles. For binary mixes with size ratio u close to unity (u ↓ 1), analytical equations for the packing increase have become available recently [31, 32]. The underlying analytical model is based on Onsager's excluded volume model from 1949 [33]. Onsager determined the orientally averaged excluded volume of two spherocylinders or two cylinders with unequal lengths and diameters. In essence this is an example of a "athermal" statistical geometric approach. In [31], this excluded volume concept of two-particle pairs was combined with the statistically probable combinations of small and large particle pairs, yielding an analytical expression for the packing fraction of binary similar particles with small size disparity. This expression contains a volume contraction function, which governs the effect of mix composition, size disparity and Euclidean dimension $\mathbb{R}^D$ on packing fraction.

This geometric approach of particle packing was successfully validated against a broad collection of computational, experimental and replica theory data of packings in three dimensions [31]. In [32], this model was applied to the packing of binary disks in plane, and to the glass transition of binary hyperspheres in the infinite spatial dimension limit. By employing integral geometry, it was evinced that the governing equations are applicable to arbitrary convex particles in $\mathbb{R}^2$ and $\mathbb{R}^3$, and hyperspheres in $\mathbb{R}^4$ [34].

For weak polydispersity, obviously the particle packing approximates the monosized packing fraction of the considered particle, which depends on its shape and the method of packing; from the loosest way ("random loose packing, RLP") to the densest way of packing ("random close packing, RCP").

This paper is organized as follows. First, in Section 2 first the model for binary particles with small size disparity is recapitulated [31, 32, 34]. For binary mixtures, the effect of polydispersity on packing fraction and glass transition depends on the size ratio u, the relative proportions of the two particle populations, and space dimension D. It was established that the dependence on these parameters is non-trivial.

Next, in Section 3, the model of the binary packing is generalized to multicomponent mixes, by considering the statistical frequencies of all particle pairs in the assembly. Next, in Section 4, packings consisting of multiple, binomially distributed, and discretely-sized particles are considered. It is furthermore shown that, when the size ratio of subsequent particle sizes is constant and tends to unity, these multiple discrete size distributions can be transformed mathematically to a continuously graded system with a lognormal distribution. Subsequently, it is demonstrated that this limit also enables an analytical expression of the packing fraction of the lognormal distribution with small standard deviation. This general closed-form equation provides the packing fraction as a function of standard deviation, the single-sized packing fraction of the particle shape considered, and the Euclidean dimension. These properties are all physically defined, without the need of an adjustable parameter. Finally, in Section 5 this original expression for the void/packing fraction is compared thoroughly with experiments and computer generated packing data concerning disks in $\mathbb{R}^2$ and a wide collection of particle shapes in $\mathbb{R}^3$, and found to be in good accordance. Next, the model is extended to the opposite limit, *viz.* infinite large standard deviations. The conclusions are collected in Section 6.

The presented models provide the random or amorphous packing of nonoverlapping (*i.e.* hard) particles in the polydisperse case relative to the monodisperse case, in D-dimensional Euclidean space $\mathbb{R}^D$. They are applicable when the assemblies are compacted equally. Whether the assembly's density corresponds to the maximally random jammed state (MRJ) [1], random close packing (RCP), random loose packing (RLP), or a configuration in between these closest or loosest possible ways of particle packing, is inconsequential. Interestingly, random particle packings are prototypical glasses in that they have no crystalline orientational or translational order, while simultaneously being mechanically rigid. Moreover, size dispersity frustrates crystallization and is therefore a glass phase enabler. Indeed the glass transition is related to a specific packing density, in "Table II" [3] packing fraction values for different protocols are listed. Also these packing fractions are affected by polydispersity and are captured by the presented model.

## 2. BINARY MIXES

In this section the assemblies of binary (discretely sized) similar particles in D-dimensional space are recapitulated, and as precursor for the case of multicomponent mixtures.

### 2.1 Distribution and volumes

A binary particle mix is considered, consisting of two components (n = 2), the larger and smaller ones with characteristic sizes $d_i$ and $d_j$ (here i = 1, j = 2), respectively, with a normalized number distribution

$$P(d) = \sum_{i=1}^{n} X_i \, \delta(d - d_i) \quad , \tag{1}$$

where δ is the Dirac delta function, and $X_i$ and $X_j$ are the number fractions of the smaller and larger components for which the following identity holds

$$\sum_{i=1}^{n} X_i = 1 \quad . \tag{2}$$

As we are considering similar particles, all dimensions of both particle dimensions have an identical size ratio $u_{i,j}$ (here i = 1 and j = 2), so that for particle volume ratio and excluded volume ratio holds

$$\frac{d_i}{d_j} = u_{i,j} \ , \ \ \frac{V_p^i}{V_p^j} = \frac{V_e^{i,i}}{V_e^{j,j}} = u_{i,j}^D = \Omega_{i,j} \quad . \tag{3}$$

As we consider similar particles, the volume ratio is the size ratio to the power of the space dimension. For spatial dimension D = 2, it constitutes the surface area ratio. Following the excluded volume model, the binary packing fraction follows from [31, 32, 34]

$$\eta = \frac{2V_p}{V_e} \quad . \tag{4}$$

with

$$2V_p = 2\sum_{i=1}^{n} X_i V_p^i \ , \qquad (5)$$

with here n = 2, as the convex addition of the two particle volumes. In Eqs. (4) and (5) the factor 2 appears as the excluded volume is based on that of the particle pair. The mean excluded volume of randomly mixing the small and large particles follows from the statistically probable combinations of these particles pairs:

$$V_e = X_i^2 V_e^{i,i} + X_j^2 V_e^{j,j} + X_i X_j (V_e^{i,j} + V_e^{j,i}) \quad . \qquad (6)$$

which is

$$V_e = X_i V_e^{i,i} + X_j V_e^{j,j} - X_i X_j (V_e^{i,i} + V_e^{j,j} - 2V_e^{i,j}) \ , \qquad (7)$$

in view of

$$V_e^{i,j} = V_e^{j,i} \ , \qquad (8)$$

and

$$X_i^2 = X_i - (1 - X_i) X_i \quad . \qquad (9)$$

Based on integral geometry [30, 33] it appeared that

$$V_e^{i,i} + V_e^{j,j} - 2V_e^{i,j} = (V_e^{j,j} - 2V_p^j) v(\Omega_{i,j}, D) \quad . \qquad (10)$$

In [31] this equation was derived based on the binary (sphero)cylinder excluded volumes provided by Onsager [33]. Using integral geometry, in both spatial dimensions D = 2 and 3, it was mathematically evinced that the excluded volume that underlies Eqs. (10) holds for arbitrary binary similar convex particles, so not for and (sphero)cylinders (D = 3) only. It also was evinced that Eq. (10) also holds for hyperspheres in spatial dimension D = 4.

### 2.2 Contraction function

In Eq. (10), the contraction function $v(\Omega_{i,j}, D)$ appears, defined as [31, 33, 34]:

$$v(\Omega_{i,j}, D) = \Psi(\Omega_{i,j}) \Delta(D) \quad , \qquad (11)$$

with

$$\Psi(\Omega_{i,j}) = \frac{2(\Omega_{i,j} - 1)^2}{(\Omega_{i,j} + 1)} \quad , \qquad (12)$$

and

$$\Delta(D) = \frac{(1 - D^{-1})}{4(1 - 2^{1-D})} \quad . \qquad (13)$$

Eq. (12) reflects that difference in volumes, and Eq. (13) accounts for the spatial dimension D [32]. In Figure 1, Δ is set out against D. The function Δ accounts for all Euclidean dimension, and Δ = ¼ for D = 2 and D → ∞, which is the maximum of Δ. For D = 3, Δ = 2/9 (≈ 0.222), which is close to the minimum value Δ ≈ 0.213 that is attained at non-Euclidean dimension D ≈ 4.72 [32].
The original contraction function is based on the excluded volume model expressions provided by Onsager for (sphero)cylinders, but it appeared that these original expression were accurate for weak polydispersity (so near u = 1). For larger size differences the above functions, which are asymptotically identical to the Onsager based expression towards u = 1, seemed to be more accurate in a larger size difference range [31, 32]. In [31, 32] it was seen that in $\mathbb{R}^2$ and $\mathbb{R}^3$ the modified model is accurate up to a volume ratio of at least 9. Furthermore, Eq. (11) is applicable to arbitrary spatial dimensions, which is not the case for the original Onsager based expressions [31, 32]. This feature was helpful in [32], where the glass transition of binary hyperspheres in the infinite spatial dimension limit was studied.

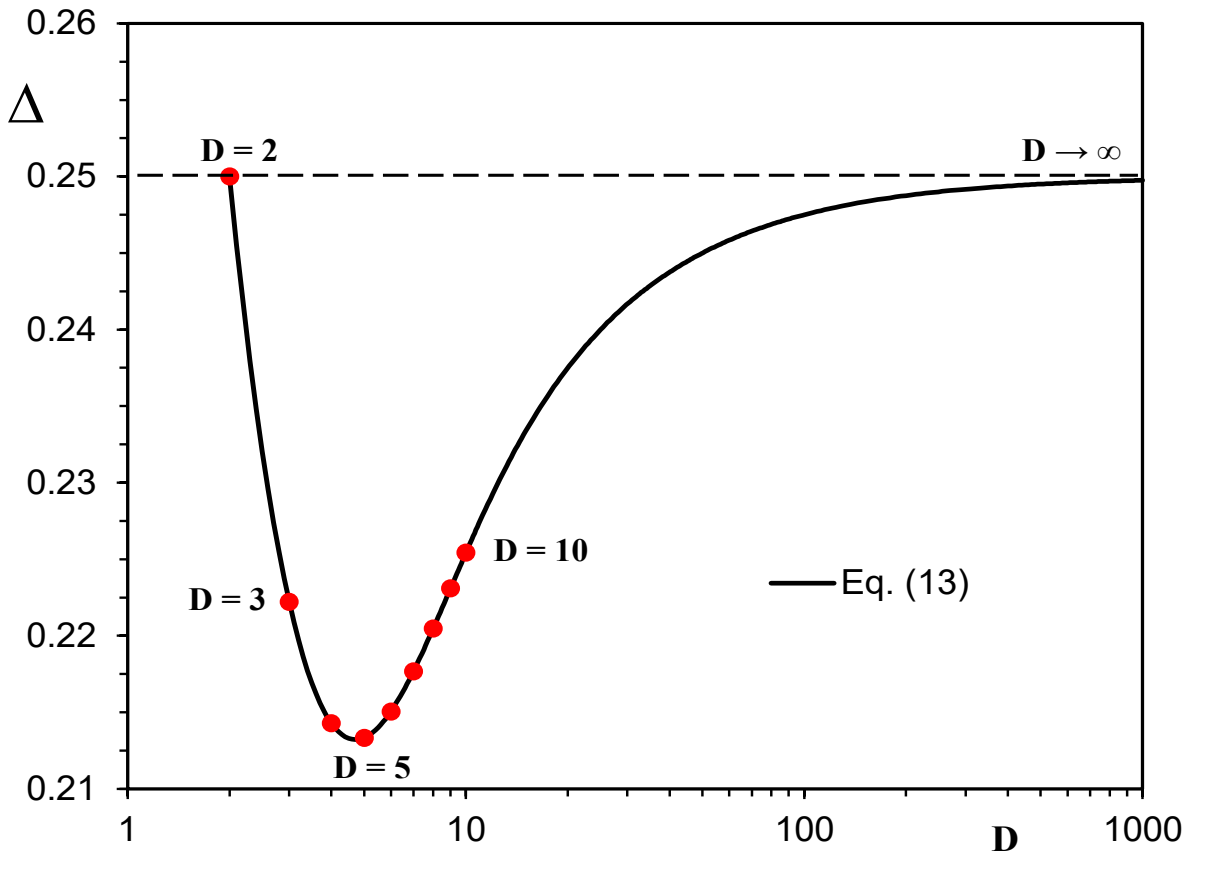

Figure 1 Function Δ(D), Eq. (13), which accounts for Euclidean dimension D on contraction function.

### 2.3 Binary packing fraction

Dividing the numerator and denominator by $V_e^{j,j}$, applying Eqs. (2) and (3), and in view that the monosized packing fraction is defined as

$$f = \frac{2V_p^j}{V_e^{j,j}} \quad , \qquad (14)$$

one obtains

$$\eta(\Omega_{i,j}, X_i, D) = \frac{f(X_i(\Omega_{i,j}-1)+1)}{X_i(\Omega_{i,j}-1)+1-X_i(1-X_i)(1-f)v(\Omega_{i,j},D)} , \quad (15)$$

with $\eta(\Omega_{i,j}, X_i, D)$ as binary packing fraction, f as monosized packing fraction, $\Omega_{i,j}$ as volume ratio (Eq. (3)), and $v(\Omega_{i,j}, D)$ as contraction function (eq. (11)), and D as the spatial dimension.
The numerator of Eq. (15) reflects the total volume of the particles, and the denominator the total volume of the packing [31, 32, 34].
Eq. (15) reveals that that the effect of bidispersity on packing fraction is governed by the product $X_i(1 - X_i)(1 - f)v(u, D)$, where $X_i(1 - X_i)$ accounts for the composition, $(1 - f)$ for the monosized void fraction (depending on particle shape and densification) and $v(\Omega_{i,j} D)$ for the contraction function, which depends on the volume ratio of the unequal particle pairs ($\Omega_{i,j}$), and the Euclidean dimension.

### 3. MULTICOMPONENT MIXES

Here, the binary mix model is extended to multi-component mixes. The mix consists of n particles with sizes $d_1, d_2, \ldots, d_n$ ($d_i > d_{i+1}$) with number fractions $X_1, X_2, \ldots, X_n$, that obey Eq. (2), and the particle size distribution is governed by Eq. (1). The convex addition of all particle pair volumes, multiplied by two, is given by Eq. (5).
The excluded of all particle pairs follows from their statistically probable combinations:

$$V_e = \sum_{i=1}^{n}\sum_{j=1}^{n} X_i X_j V_e^{i,j} = \sum_{i=1}^{n} X_i^2 V_e^{i,i} + 2\sum_{i=1}^{n-1}\sum_{j=i+1}^{n} X_i X_j V_e^{i,j} . \quad (16)$$

In view of Eq. (2) we may write

$$X_i^2 = X_i\Big(1-\sum_{j=1}^{n} X_j + X_i\Big) , \quad (17)$$

so that Eq. (16) becomes

$$V_e = \sum_{i=1}^{n} X_i V_e^{i,i} + \sum_{i=1}^{n} X_i\Big(X_i - \sum_{j=1}^{n} X_j\Big)V_e^{i,i} + 2\sum_{i=1}^{n-1}\sum_{j=i+1}^{n} X_i X_j V_e^{i,j} . \quad (18)$$

It furthermore follows that

$$\sum_{i=1}^{n} X_i\Big(X_i - \sum_{j=1}^{n} X_j\Big)V_e^{i,i} = -\sum_{i=1}^{n-1}\sum_{j=i+1}^{n} X_i X_j (V_e^{i,i} + V_e^{j,j}) , \quad (19)$$

and hence Eq. (18) can be written as

$$V_e = \sum_{i=1}^{n} X_i V_e^{i,i} + \sum_{i=1}^{n-1}\sum_{j=i+1}^{n} X_i X_j (2V_e^{i,j} - V_e^{i,i} - V_e^{j,j}) . \quad (20)$$

In view of Eq. (10), the last term can be written as:

$$\sum_{i=1}^{n-1}\sum_{j=i+1}^{n} X_i X_j (2V_e^{i,j} - V_e^{i,i} - V_e^{j,j}) = -\sum_{i=1}^{n-1}\sum_{j=i+1}^{n} X_i X_j (V_e^{j,j} - 2V_p^{j}) v(\Omega_{i,j}, D) . \quad (21)$$

As relative volumes are relevant, the particle volumes and the excluded volumes are both divided by the smallest monosized excluded volume, *viz.* $V_e^{n,n}$. Hence, Eq. (5) and the first term of Eq. (18) become:

$$\frac{2V_p}{V_e^{n,n}} = 2\sum_{i=1}^{n} \frac{X_i V_p^i V_e^{i,i}}{V_e^{i,i} V_e^{n,n}} = f\sum_{i=1}^{n} X_i \Omega_{i,n} , \quad (22)$$

and

$$\frac{V_e}{V_e^{n,n}} = \sum_{i=1}^{n} X_i \Omega_{i,n} , \quad (23)$$

respectively, see Eqs. (3) and (14). The last term of Eq. (21) can be written as

$$\frac{1}{V_e^{n,n}}\sum_{i=1}^{n-1}\sum_{j=i+1}^{n} X_i X_j (V_e^{j,j} - 2V_p^{j}) v(\Omega_{i,j}, D) = (1-f)\sum_{i=1}^{n-1}\sum_{j=i+1}^{n} X_i X_j \Omega_{j,n} v(\Omega_{i,j}, D) , \quad (24)$$

see Eqs. (3) and (14). Substituting Eqs. (22)-(24) in Eq. (4) yields as polydisperse packing fraction

$$\eta = \frac{f\sum_{i=1}^{n} X_i \Omega_{i,n}}{\sum_{i=1}^{n} X_i \Omega_{i,n} - (1-f)\sum_{i=1}^{n-1}\sum_{j=i+1}^{n} X_i X_j \Omega_{j,n} v(\Omega_{i,j}, D)} \quad . \quad (25)$$

In [30] the following the following normalized binary packing fraction was introduced

$$\lambda = \frac{\eta - f}{\eta(1 - f)} \quad , \qquad (26)$$

which enabled a comparison of binary packing fractions with different monosized packing fraction. Binary packing fractions of packings with different particle shapes and mode of packing collapse when they have the same distribution. By computing λ the effect of the monosized packing fraction is omitted, as λ depends on size ratio and composition (that both determine the polydispersity). For instance the RLP and RCP packing fractions of binary spheres with small size ratio indeed collapse when normalized by Eq. (26) [31], though these RLP and RCP assemblies have a very distinct monosized packing fractions f.
When Eq. (25) is substituted in Eq. (26) we have

$$\lambda = \frac{\sum_{i=1}^{n-1}\sum_{j=i+1}^{n} X_i X_j \Omega_{j,n} v(\Omega_{i,j}, D)}{\sum_{i=1}^{n} X_i \Omega_{i,n}} \quad . \qquad (27)$$

As with binary packings, also for polydisperse packings, λ is governed by the particle size distribution only, and not on particle shape and their mode of packing.

## 4. BINOMIAL AND LOGNORMAL DISTRIBUTIONS

In this section, the packing of multiple discretely sized particles, with a large size ratio, is studied for the special cases whereby their number fractions obey a binomial and lognormal distributions. These lognormal distributions are widespread in the physical, biological, social and economic sciences [35, 36]. Particles that result from natural weathering or industrial comminution [37], or are formed by condensation or solidification processes [38], often closely fit the lognormal distribution.

### 4.1 Binomial and lognormal distributions

For the binomial distribution of n particle sizes and probability p, their number fractions for i = 1, 2,…, n, are given as

$$X_i = \binom{n-1}{i-1} p^{n-i}(1-p)^{i-1} \quad . \qquad (28)$$

Henceforth, in this paper only symmetrical distributions are considered, so p = ½. It is furthermore invoked that all subsequent particle sizes, and hence also their volumes, have a constant ratio, whereby

$$\frac{d_i}{d_{i+1}} = u \quad ; \quad \frac{V_p^i}{V_p^{i+1}} = \Omega \quad ,$$

(29)

so that

$$\Omega_{i,j} = \frac{V_p^i}{V_p^j} = \Omega^{j-i} = u^{D(j-i)} \quad , \qquad (30)$$

and that hence for this symmetrical binomial distribution holds

$$Ln^2\sigma_g = \tfrac{1}{4}(n-1)Ln^2 u = \frac{\tfrac{1}{4}(n-1)Ln^2\Omega}{D^2} \quad , \qquad (31)$$

with $Ln\sigma_g$ as the standard deviation.
Adding more monosized particle sizes, all adjacent sizes having the same size ratio u, and letting u → 1, following the central limit theorem the binomial distribution turns into a continuous lognormal distribution [39]. Here the logarithm of the particle sizes are namely binomially or normally distributed, which means that a lognormal distribution of the particle size is obtained for u → 1, for a lognormal distribution is a continuous probability distribution of a random variable whose logarithm is normally distributed:

$$Lnd = N(Lnd_g, Ln^2\sigma_g) \quad , \qquad (32)$$

with $Lnd_g$ as geometric mean size.

### 4.2 Applicability of model

The presented analytical model is valid for particles which size ratio is about 1/9 ≤ Ω ≤ 9. The lognormal size distribution stretches from d = 0 to ∞, so that this criterion is not met for all particles, especially for larger standard deviations. It is possible to assess the fraction of particles that is located within the model's range, that is the probability of two random particles, with sizes $d_1$ and $d_2$, having a volume ratio Ω that meets 1/9 ≤ Ω ≤ 9, so the probability that for two arbitrary sizes

$$\frac{d_1}{u_m} \le d_2 \le u_m d_1 \quad , \qquad (33)$$

or

$$- Lnu_m \leq Lnu \leq Lnu_m \quad , \tag{34}$$

with

$$u_m = {}^{D}\log 9 \quad . \tag{35}$$

As

$$Lnd_1, Lnd_2 = N(Lnd_g, Ln^2\sigma_g) \quad , \tag{36}$$

and

$$Lnu = Lnd_1 - Lnd_2 \quad , \tag{37}$$

it follows that

$$Lnu = N(0, 2Ln^2\sigma_g) \quad , \tag{38}$$

as the difference of the geometric means of the $Lnd_1$ and the $Lnd_2$ distributions is zero, and since $Lnd_1$ and $Lnd_2$ are distributed independently, their variances are summed [39]. Accordingly, the probability that two sizes are located within the model range, reads

$$P_m = P\left(- Lnu_m \leq Lnu \leq Lnu_m\right) = 2\Phi\left(\frac{Lnu_m}{Ln\sigma_g\sqrt{2}}\right) - 1 \quad , \tag{39}$$

with $\Phi$ as cumulative distribution function (CDF) of the standard normal distribution.

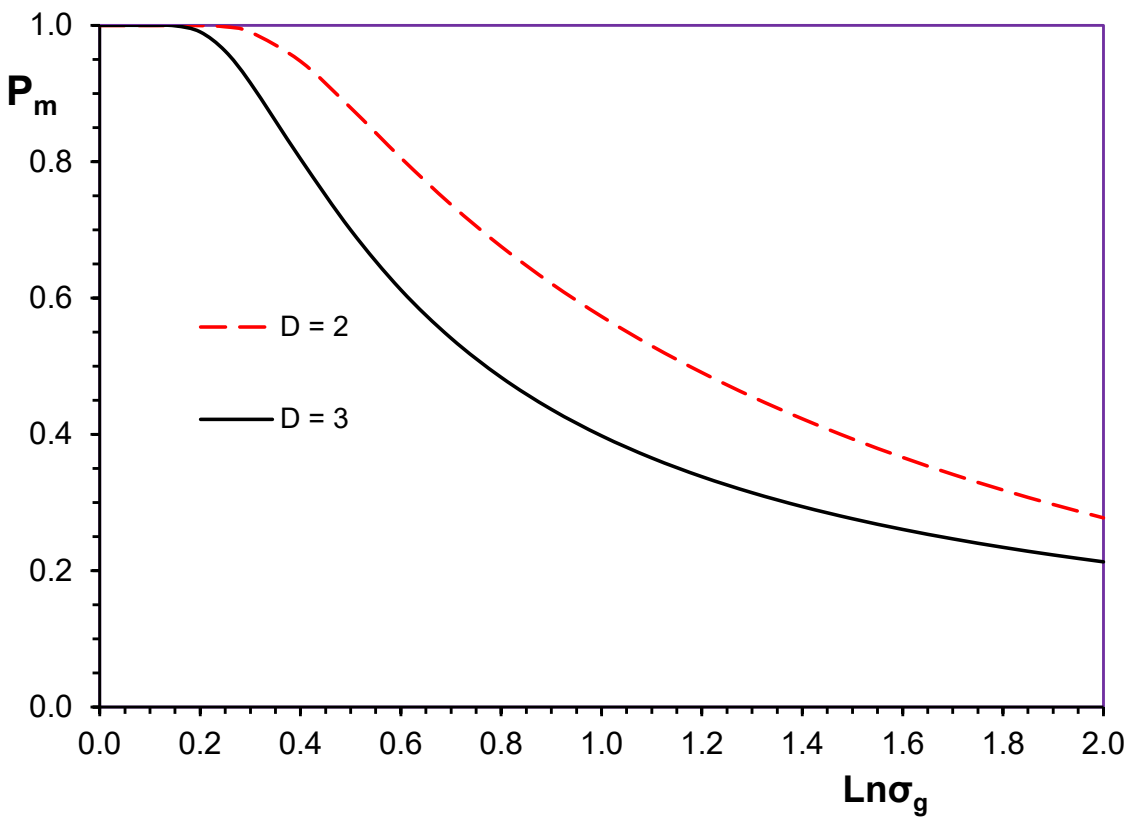

Figure 2 Fraction of particles $P_m$ of a lognormal size distribution that have a volume ratio smaller than 9 versus their standard deviation $Ln\sigma_g$ for space dimensions D = 2 and D = 3 (Eqs. (35) and (39)).

In Figure 2, $P_m$ is set out versus $Ln\sigma_g$ for $u_m = 2.08$ and $u_m = 3$, the former value pertains to Euclidean dimension D = 3, the latter to D = 2, see Eq. (35). One can see, as expected, that with increasing $Ln\sigma_g$, so with wider distribution, a smaller fraction of the particles fall within the model range. Notwithstanding this limitation, it is interesting to apply the multicomponent model to the lognormal distribution.

### 4.3 Particles volume

In this subsection the convex addition of the volumes of all particles that are lognormally distributed is computed, that appears both in numerator and denominator of the packing fraction, Eq. (25), and in the denominator of the scaled packing fraction (Eq. (27)).
Eqs. (28) and (30) yield

$$\sum_{i=1}^{n} X_i \Omega_{i,n} = \sum_{i=1}^{n} X_i \Omega^{n-i} = \left(\tfrac{1}{2}\right)^{n-1} \sum_{i=0}^{n-1} \binom{n-1}{i} \Omega^{n-i-1} = \left(\tfrac{1}{2}\right)^{n-1} \left(1+\Omega\right)^{n-1} \quad . \tag{40}$$

Substituting the following approximation

$$\Omega = e^{\frac{1}{\varepsilon}Ln\Omega^{\varepsilon}} = 1 + \frac{Ln\Omega^{\varepsilon}}{\varepsilon} + \frac{Ln^2\Omega^{\varepsilon}}{2\varepsilon^2} + O\left(\varepsilon^3\right) \quad , \tag{41}$$

in Eq. (40) and letting $\varepsilon = (n-1)^{-1} \downarrow 0$, and applying Eq. (31), yields

$$\sum_{i=1}^{n} X_i \Omega^{n-i} = \Omega^{(n-1)/2} e^{\frac{(n-1)Ln^2\Omega}{4}} = \Omega^{(n-1)/2} e^{Ln^2\sigma_g^D} \quad . \tag{42}$$

### 4.4 Alternative contraction function

The combination of Eqs. (11)-(13) and (25) enable the computation of the contraction term in the packing fraction of multicomponent mixes, *i.e.* second term in the denominator of Eq. (25), for a given monosized packing fraction, space dimension, composition and particle volume distribution.

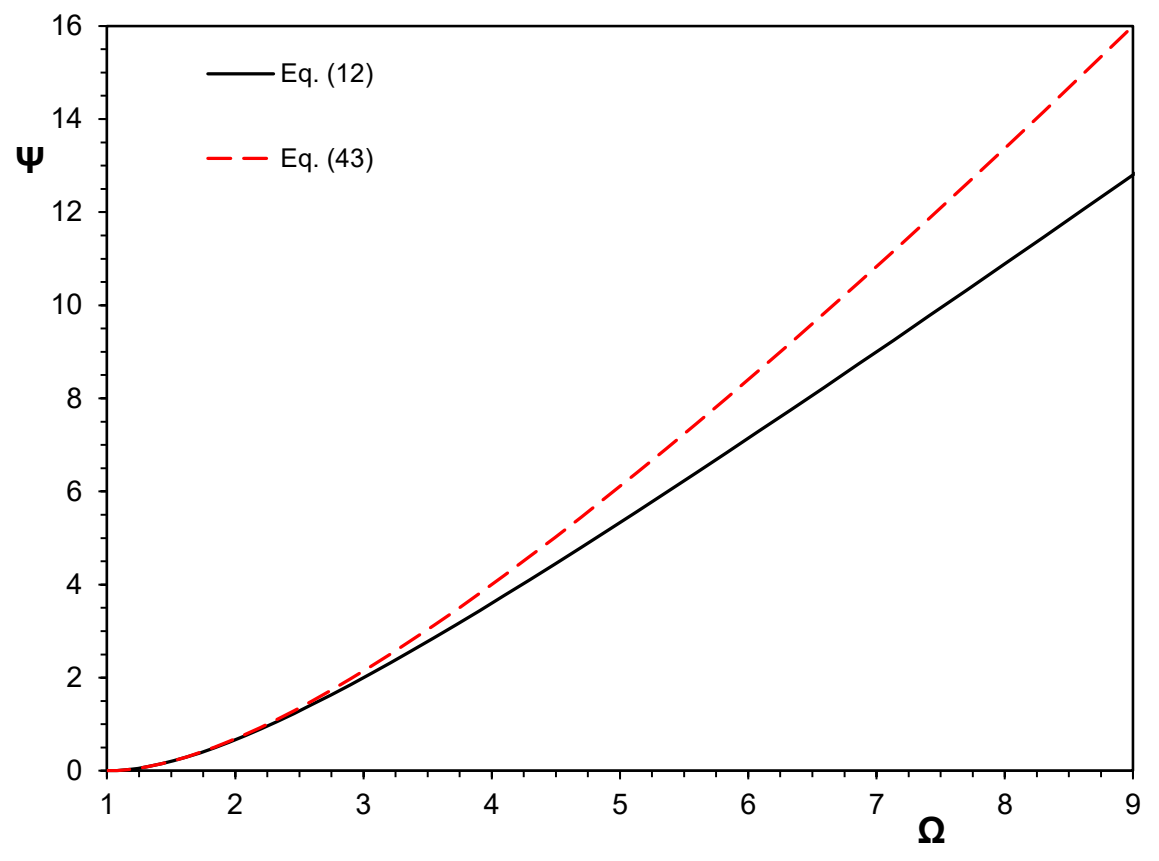

Figure 3 Contraction function $\Psi$ for a particle pair according to Eq. (12), and its approximation Eq. (43), versus their volume ratio $\Omega_{i,j}$.

To obtain a system that is more tractable analytically (next subsection), we introduce an alternative expression for Eq. (12), which is asymptotically identical towards $\Omega_{i,j}$ being unity (particle volumes of particles i and j becoming equal), and sufficiently accurate in the range $1 \le \Omega_{i,j} \le 9$. To this end, the following approximation is introduced

$$\Psi(\Omega_{i,j}) = 4(\Omega_{i,j}^{1/2} - 1)^2 \quad . \tag{43}$$

In Figure 3, Eq. (12) and this approximation are shown. Towards $\Omega_{i,j} = 1$, both equations are asymptotically identical in $\Omega_{i,j}$.
Furthermore, Eq. (12) stems from $(V_i - V_j)^2/((V_i + V_j)$ [31], which constitutes a volume, in dimensional form Eq. (40) corresponds to $(\sqrt{V_i} - \sqrt{V_j})^2$, which is also a volume. So also from physical point of view, Eq. (43) has the right characteristics.

### 4.5 Contraction term

Next, the contraction term, this is the last term in the denominator of the packing fraction (Eq. (25)), and the numerator of the scaled packing fraction (Eq. (27)), is determined. Using Eqs. (11), (28), (30) and (43):

$$\sum_{i=1}^{n-1}\sum_{j=i+1}^{n} X_i X_j \Omega_{j,n} v(\Omega_{i,j}, D) =$$

$$4\Delta \sum_{i=1}^{n-1}\sum_{j=i+1}^{n} X_i X_j (\Omega^{(n-i)/2} - \Omega^{(n-j)/2})^2 = \tag{44}$$

$$2\left(\tfrac{1}{2}\right)^{2n-2} \Delta \sum_{i=0}^{n-1}\sum_{j=0}^{n-1} \binom{n-1}{i}\binom{n-1}{j} (\Omega^{n-i-1} - 2\Omega^{(2n-j-i-2)/2} + \Omega^{n-j-1}) \quad .$$

In this transition use has been from the symmetry in the summation: for i = j the sum of the three terms is zero, and sums of (i, j) and (j, i) are identical.
To the first (and last) terms of the summation, Eqs. (40)-(42) apply. Using Eq. (41) with $\Omega$ replaced by $\sqrt{\Omega}$, the middle term of Eq. (44) becomes

$$\left(\tfrac{1}{2}\right)^{2n-2} \sum_{i=0}^{n-1}\sum_{j=0}^{n-1} \binom{n-1}{i}\binom{n-1}{j} \Omega^{(2n-i-j-2)/2} =$$

$$\left( \left(\tfrac{1}{2}\right)^{n-1} \sum_{i=0}^{n-1} \binom{n-1}{i} \left(1+\sqrt{\Omega}\right)^{n-i-1} \right)^2 = \tag{45}$$

$$\left( \left(\tfrac{1}{2}\right)^{n-1} \left(1+\sqrt{\Omega}\right)^{n-1} \right)^2 = \Omega^{(n-1)/2} e^{\frac{(n-1)Ln^2\Omega}{8}} \quad ,$$

for $n - 1 \to \infty$. With Eq. (42) it furthermore follows:

$$\Omega^{(n-1)/2} e^{\frac{(n-1)Ln^2\Omega}{8}} = \Omega^{(n-1)/2} e^{\frac{Ln^2\sigma_g^D}{2}} \quad . \tag{46}$$

### 4.5 Packing fraction

Combining Eqs. (42), (44) and (46), Eq. (25) yields as lognormal packing fraction

$$\eta = \frac{f}{1 - 4(1-f)\left(1 - e^{\frac{-Ln^2\sigma_g^D}{2}}\right)\Delta(D)} \quad . \tag{47}$$

The scaled faction fraction follows from Eqs. (26) and (47) as

$$\lambda = 4\left(1 - e^{\frac{-Ln^2\sigma_g^D}{2}}\right)\Delta(D) \quad . \tag{48}$$

In [31] this transformation was introduced for the binary packing fraction of similar particles in space dimension D = 3 to eliminate the influence of the monosized packing fraction. For the lognormal distribution we see that, by introducing $\lambda$, also here an expression is obtained that is governed by the polydispersity (solely determined by $\log\sigma_g$) and the dimension D.
In Eq. (48) the effect of Euclidean dimension D becomes evident, by the factor $\Delta(D)$ and by $DLn\sigma_g$ in the exponential function. Note that $4\Delta$ is unity for D = 2 and $D \to \infty$, and close to unity for intermediate Euclidean dimensions. The maximum deviation from unity takes place for D = 4 and 5, for which $4\Delta$ = 6/7 and 64/75, respectively, see Eq. (13) and Figure 1.

## 5. COMPARISON WITH REPORTED DATA

In the section the model predictions are compared with a large collection of experimental data, simulations and theoretical approaches reported before. All data can be found in the Appendix.

### 5.1 Reported data

For continuous normal and lognormal distributions of sands, Sohn and Moreland [13] determined experimentally the packing fraction of sands (MR and OS), as a function of the standard deviation (Tables A.1-A.4). For discrete lognormal distributions of steel spheres Dexter and Tanner [14] measured the packing fraction as function of the standard deviation (Table A.5). Suzuki et al. [16] measured the packing fraction of broken glass and glass beads, that were compacted by tapping or prodding (Tables A.6-A.9). Yang et al. [17], He et al. [18], and Farr and Groot [19] reported computational simulations of sphere packings with different compactions (Tables A.10-A.13). Baranau and Tallarek [22] modelled the effect of lognormal polydispersity on packings with "lowest possible density", "densest possible density" and "densest glass close packing limit" (Tables A.14-A.16). Malmir et al. [24] simulated the packing of cubes (Tables A.17). Desmond [40] simulated the packing of spheres using both the algorithms of [25] and Farr and Groot [19] (Tables A.18 and A.19). The aforementioned data all concerns packings of particles in $\mathbb{R}^3$. For packing of disks with a lognormal size distribution in a plane, so in two dimensions, numerically simulated packing fractions were provided by Desmond [40] and using the algorithm of [25] (Table A.20), and by Weeks [41], using the algorithm of [42] (Table A.21).
The particles possess a monosized packing fraction that depends on particle shape, which varies, and the method of packing (between RLP and RCP), which is also the case for the lognormal distributions composed of them. In order compare all of them, the scaled packing fractions as given by Eq. (26) are used (Appendix), and compared with Eq. (48).

### 5.2 Weak polydispersity

First, Eq. (48) is compared with packing data of particles in $\mathbb{R}^3$ and with disks in $\mathbb{R}^2$ (Figure 4), for weak polydispersity, so in the vicinity of $\mathrm{Ln}\sigma_g = 0$, for which Eqs. (47) and (48) are applicable. To eliminate the effect of space dimension D, $\lambda/\Delta$ is set out versus $\mathrm{Dln}\sigma_g$, see Eq. (48). For D = 2 and D = 3, $\Delta = 1/4$ and $2/9$, respectively (Figure 1). In Figure 4 also reported data on a collection of different particle shapes and packing modes, are included (D = 3), and disk packings (D = 2), taken from the Appendix.

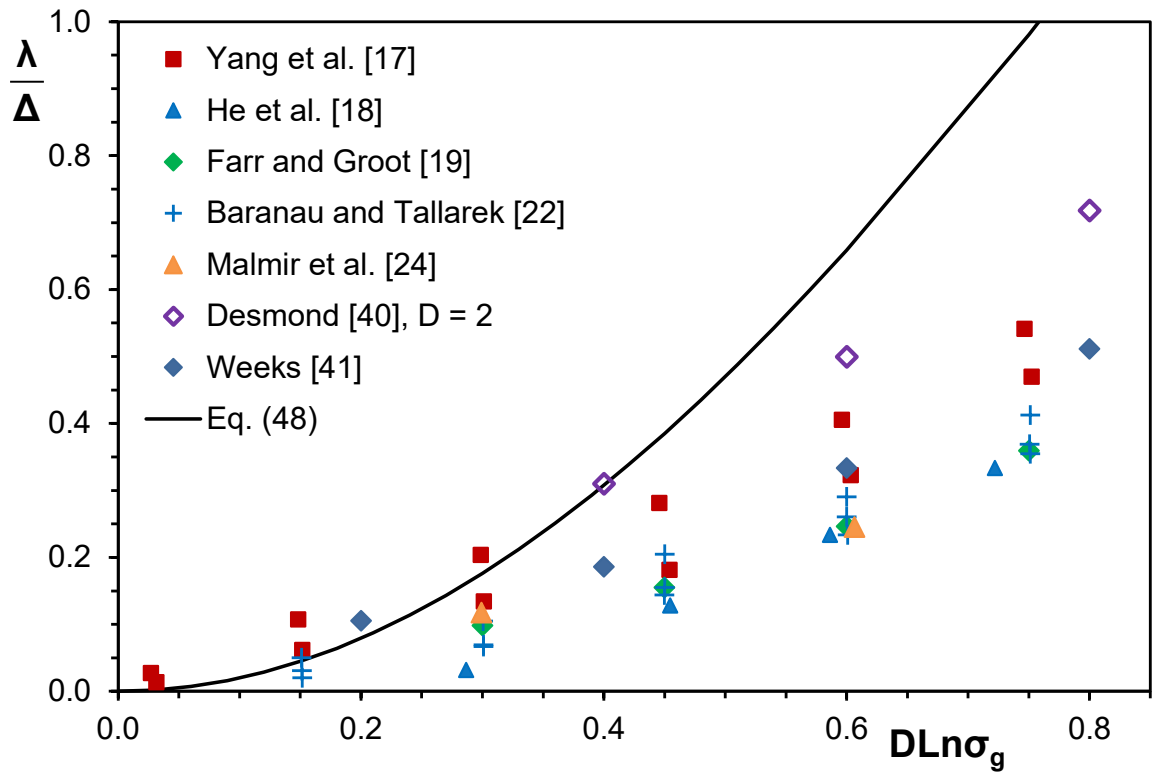

Figure 4 Scaled packing fraction of lognormal particle distributions, $\lambda/\Delta$, following Eq, (48), and packing data taken from literature (Appendix), versus product of Euclidean dimension and standard deviation, $\mathrm{DLn}\sigma_g$, from 0 to 0.85.

The relevant scaling factor (Eq. (26)), $1 - f$, which is the monosized void fraction, ranges from about 0.16 for disks in D = 2, to larger than 0.35 for the particles in D = 3 (Appendix). Though these $1 - f$, the space dimensions D, and the dimension function $\Delta(D)$ (Figure 1), all vary greatly, we can see that the scaled values all follow the same trend and are lying close to each other. One can furthermore see that the model, from which the scaling is derived, follows all data very closely for small $\mathrm{DLn}\sigma_g$. For larger $\mathrm{DLn}\sigma_g$, the model overshoots the effect of polydispersity on packing fraction. This is due to a) an approximation of the contraction function is used that is slightly larger than the original (Figure 3), the original contraction function is valid for small size disparities only, which is less the case for larger $\mathrm{DLn}\sigma_g$ (Figure 2).

### 5.2 Large polydispersity

In Figure 5, scaled packing data up to much larger standard deviation $\mathrm{Ln}\sigma_g$ are shown (Appendix). Though the variation in particle shapes and packing modes for the three dimensional particles are very different, their scaled packing fractions are lying together and following the same trend.

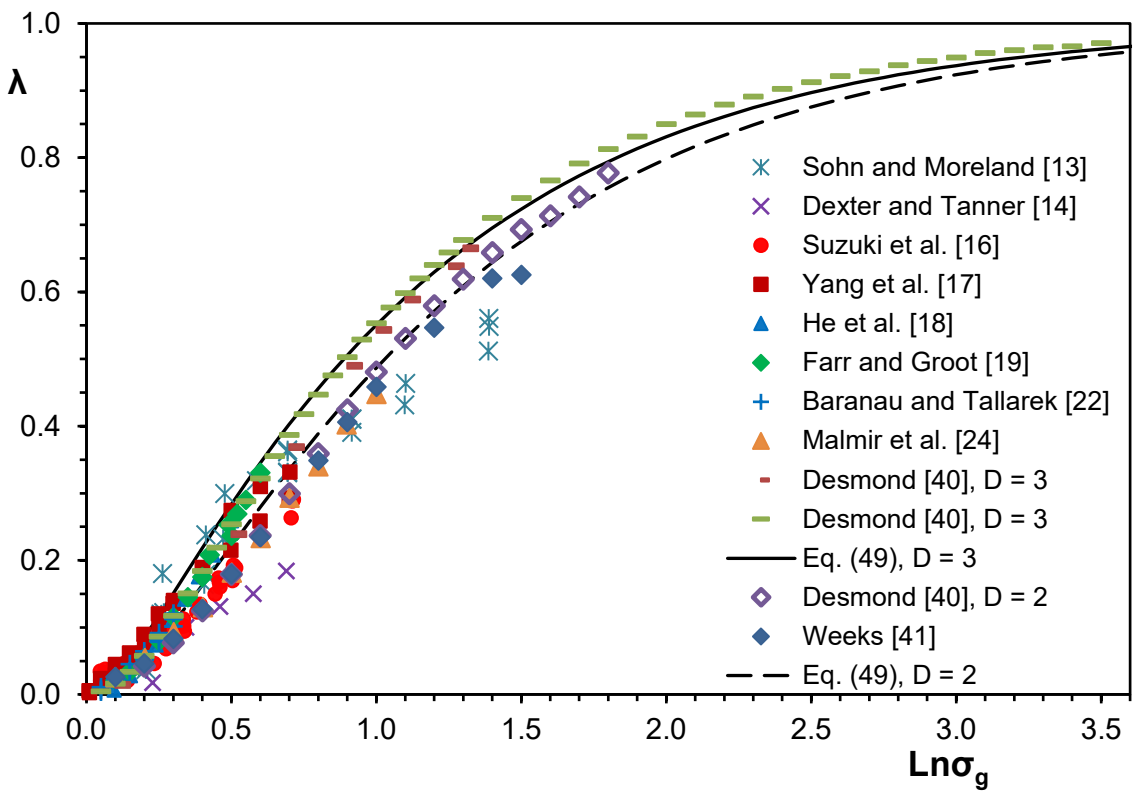

Figure 5 Scaled packing fraction of lognormal particle distributions, λ, following Eq, (48) for D = 2 and D = 3, and packing data taken from literature (Appendix), versus the standard deviation $\mathrm{Ln}\sigma_g$ from 0 to 3.6.

In this figure also the simulation results for disks in a plane are included again. In Figure 4 we have seen when the scaled packing fractions are set out against $\mathrm{DLn}\sigma_g$, for small $\mathrm{DLn}\sigma_g$, the data from $\mathbb{R}^3$ and $\mathbb{R}^2$ collapse, in line with the prediction by Eq. (48).
On the other hand, for large $\mathrm{Ln}\sigma_g$, it can be seen that the scaled packing fractions of particles in $\mathbb{R}^2$ and $\mathbb{R}^3$ coincide when set out against $\mathrm{Ln}\sigma_g$, *i.e.* for large $\mathrm{Ln}\sigma_g$ the scaled packing fraction is a unique function of $\mathrm{Ln}\sigma_g$, regardless of Euclidean dimension D. Furthermore, in [13] it was asserted that the lognormal packing fraction reaches a maximum, approached asymptotically by $1/\sigma_g$, which is compatible with $\exp(-\mathrm{Ln}\sigma_g)$.

### 5.3 Arbitrary polydispersity

Based on the analytical solution for $\mathrm{Ln}\sigma_g$, the scaled packing fractions depicted in Figures 4 and 5, and the asymptotic behaviour invoked by [13], the following approximate expression is proposed for the scaled packing fractions:

$$\lambda = 1 - e^{\frac{-2\mathrm{Ln}^2\Delta\sigma_g^D}{1+2D\Delta\mathrm{Ln}\sigma_g^D}} \quad . \tag{49}$$

For $\mathrm{Ln}\sigma_g$ close to zero, this expression approaches Eq.(48), whereas for large $\mathrm{Ln}\sigma_g$ it approximates the asymptotic values $\lambda = 1$ by $1/\sigma_g$, regardless the space dimension D. This limit $\lambda = 1$ for infinite standard deviation corresponds to a packing fraction of unity, see Eq. (26).

In Figure 5, Eq. (49) has been included using D = 2 and D = 3. Figure 5 shows that the proposed fit follows the data from literature very well for large $\mathrm{Ln}\sigma_g$, and for D = 2 and 3 the packing fractions then converge to the same asymptotic limit. Indeed Figure 5 shows that for large $\mathrm{Ln}\sigma_g$ the distinction between space dimensions disappears and the expression is able to predict the scale packing fraction well for the D = 2 and D = 3 data, that then start to coincide.

It is also to noteworthy to see how this approximation matches with the analytical solution derived for small $D\mathrm{Ln}\sigma_g$, viz. Eq. (48). To this end, in Figures 6 and 7 the scaled packing fraction $\lambda$ is set out against $\mathrm{Ln}\sigma_g$ for D = 2 and 3, respectively. Again also the data from literature is shown (Appendix), and Eqs. (48) and (49).

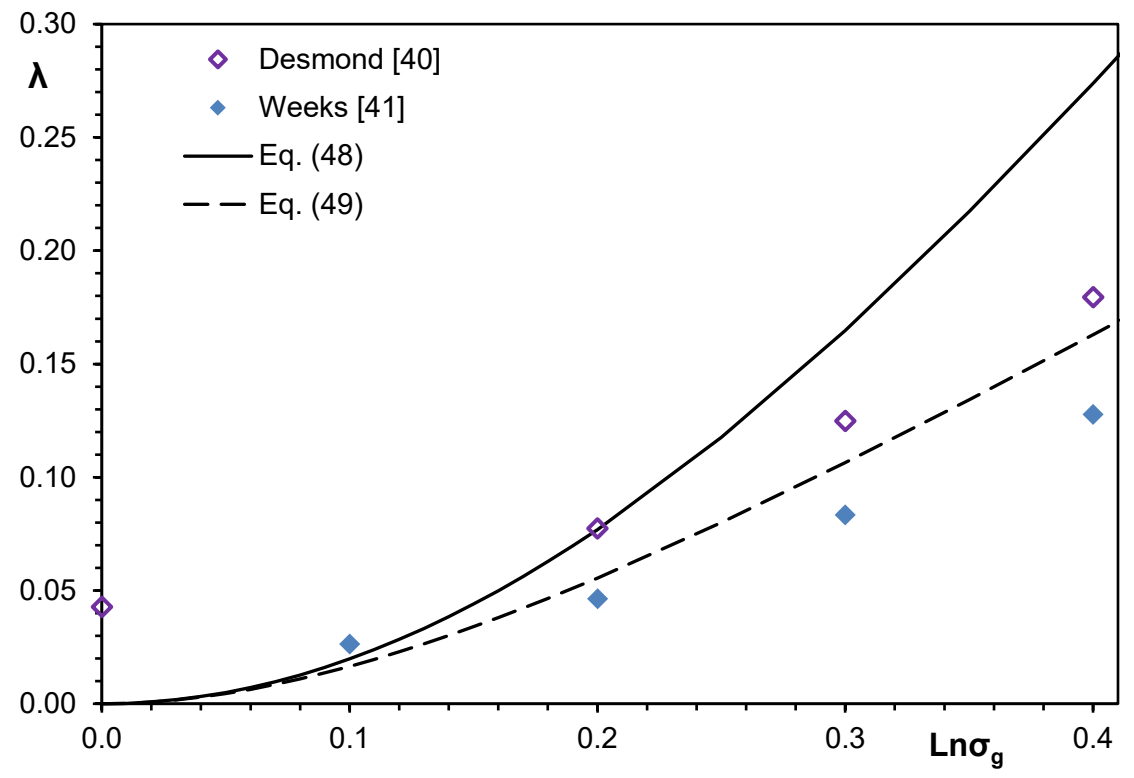

Figure 6 Scaled packing fraction of lognormal particle distributions, $\lambda$, following Eqs, (48) and (49) for D = 2, and results of packing data (Appendix), versus standard deviation $\mathrm{Ln}\sigma_g$ from 0 to 0.41.

The figures show that the effect of polydispersity is smaller in space dimension D = 2 than for D = 3, for equal standard deviation. This trend is in line with Eq. (48), which predicts that the effect of polydispersity is proportional to $(D\mathrm{Ln}\sigma_g)^2$ near $\mathrm{Ln}\sigma_g = 0$.

Both figures again confirm that $\lambda$ following Eq, (48), set out for D = 2 and D = 3, for $\mathrm{Ln}\sigma_g$ from 0 to 0.4, follows all reported data quite well close to $\mathrm{Ln}\sigma_g = 0$, which was also seen in Figure 4.

One can furthermore see that approximate Eq. (49) matches the reported data very well for larger $\mathrm{Ln}\sigma_g$ too. In other words, both for small and large $\mathrm{Ln}\sigma_g$, it provides an accurate prediction of the packing fraction.

To express the obtained results in packing fraction, the inverse of Eq. (26) is employed:

$$\eta = \frac{f}{1 - \lambda(1 - f)} \quad . \tag{50}$$

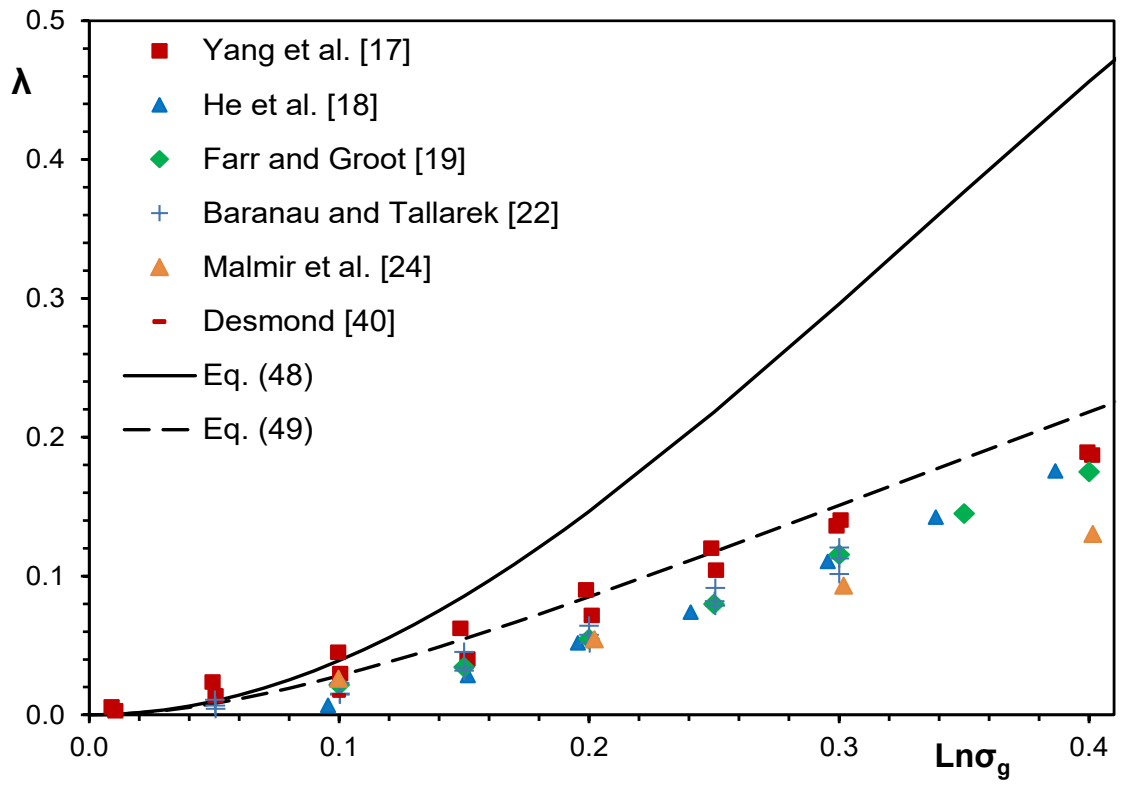

Figure 7 Scaled packing fraction of lognormal particle distributions, $\lambda$, following Eqs. (48) and (49) for D = 3, and results of packing data (Appendix), versus standard deviation $\mathrm{Ln}\sigma_g$ from 0 to 0.41.

Substituting Eq. (49) yields as lognormal packing fraction

$$\eta(\mathrm{Ln}\sigma_d, D) = \frac{f}{1 - (1 - f)\left(1 - e^{\frac{-2D^2\Delta\mathrm{Ln}^2\sigma_g}{1+2D^2\Delta\mathrm{Ln}\sigma_g}}\right)} \quad . \tag{51}$$

Eq. (51) matches the analytical model for small $\mathrm{Ln}\sigma_g$ (Eq. (47)), but is applicable to arbitrary space dimension D, and the asserted asymptote for large $\mathrm{Ln}\sigma_g$, and is validated for Euclidean dimensions D = 2 and 3.

## 6. CONCLUSION

This paper addresses the packing fraction of multi-sized similar hyperparticles, with application to binomial and a lognormal size distributions. A closed-form equation is derived for the packing fraction of multi-sized particles using a statistical approach, based on the volume contraction when particles of two different sizes are combined. Here, the bidisperse packing forms the basis of the packing description of polydisperse arrangements. Accordingly, first, the packing of discretely sized binary hyperparticles with small polydispersity is studied. The model of [31, 32], derived by combining Onsager's excluded volume model of particle pairs [33], and their statistical occurrence, is recapitulated. This excluded volume based model, governed by Eqs. (11) and (15) was validated for packing of binary disks in $\mathbb{R}^2$, for particles in $\mathbb{R}^3$, and for the glass transition of hyperspheres in the infinitely large-dimension limit [31, 32].

Next, this binary particle model is extended to an assembly of n particles, the effect of particle packing governed by the statical occurrence of the possible particle pairs, resulting in Eq. (25). Hence, a statistical approach is developed to the packing of polydisperse similar particles with small size difference, that is formulated directly in terms of the particle size distribution. The formulation is not restricted to a particular distribution function and can therefore accommodate discrete, continuous, multimodal, and skewed particle-size distributions. The central objective is to establish how the statistical population of particle sizes translates into the available volume and, ultimately, into the packing fraction of the assembly. The binomial and lognormal distributions, as particular analytical distributions, are considered as examples rather than as prerequisite of the theory.

Consequently, first Eq. (25) is applied to particles with a binomial distribution and having a constant volume ratio Ω between each subsequent size. Taking an infinite number of particle sizes, and letting the size ratio u between two subsequent sizes approach unity, a continuous lognormal distribution is obtained.

To enable a closed-form solution, an approximate contraction function is introduced, Eq. (43), which is asymptotically identical to the original excluded volume-based expression in the small volume difference limit, and in good agreement with the original in its validity range (Figure 2).

With Eq. (43) the packing fraction can be solved analytically, the resulting expression, Eqs. (47) and (48), valid for arbitrary Euclidean dimensions D when the standard deviation $\mathrm{Ln}\sigma_g$ is sufficiently close to zero (Figure 3). In Figure 4 the scaled expression, Eq. (48), is shown for particles in D = 2 and 3, and is compared with data for literature, yielding good agreement.

Subsequently, the opposite limit, large standard deviations, is addressed. Packing data in from literature, scaled according to Eq. (26), provide sufficient evidence that in $\mathbb{R}^2$ and $\mathbb{R}^3$ the packing fraction asymptotically tends to the same maximum, indicating that this limit does not depend on Euclidean dimension D. This limit represents the maximum attainable random packing fraction of particles with a lognormal size distribution. In line with [13], it is asserted that this asymptote is approximated by $1/\sigma_g$ for large $\sigma_g$.

The study reveals that for small $\mathrm{Ln}\sigma_g$, the effect of polydispersity on packing fraction (and glass transition) is governed by $D\mathrm{Ln}\sigma_g$, whereas for large $D\mathrm{Ln}\sigma_g$, it is governed by $\mathrm{Ln}\sigma_g$ . Hence, an original expression for the scaled packing fraction is put forward that matches the closed-form solution for small $D\mathrm{Ln}\sigma_g$, and the observed asymptotic behavior for large $D\mathrm{Ln}\sigma_g$ (Eqs. (49) and (51)). Figures 4-7 confirm that this expression captures the scaled packing fraction very well in the entire $\mathrm{Ln}\sigma_g$ range for both D = 2 and 3.

Interestingly, for binary hypersphere packings the effect of small polydispersity was governed by $u^D$ (volume ratio of the two particles), whereas for large u the space dimension did not play a role anymore either [31]. In other words, for large and small polydispersity, binary and lognormal distributions feature the same characteristics with respect to the Euclidean dimension.

The packing fraction of the lognormal distribution, depends on the compaction, which are asserted all to be identical, so from $\mathrm{Ln}\sigma_g = 0$, which yields the monosized packing fraction f, and beyond. This compaction may correspond to MRJ, RCP, RLP, glass (transition) density *etc.*. The present model for the effect of lognormal polydispersity on packing fraction (and glass transition) can cope with all states of densification. The governing closed-form expression contain the monosized packing fraction f, the standard deviation $\mathrm{Ln}\sigma_g$, and the Euclidean dimension D only.

## ACKNOWLEDGMENTS

Dr. K.W. Desmond is acknowledged for providing simulations on disk and sphere packings (Tables A.18-A.20), and Dr. E. Weeks (Emory University) for providing computer simulations of binary disk packings (Table A.21), and Dr. S. Spiez and Dr. G. Hüsken are thanked for extracting the data of Refs. [13], [14], [17] and [19], and Dr. Y. Luo and Mrs. Y.L. Chen for extracting the data of Refs. [16], [19], [22] and [24], see Appendix.

## APPENDIX

In this Appendix measured and simulated packing fractions of lognormal distributions are recapitulated.

| $\log\sigma_g$ | η | λ |
|---|---|---|
| 0 | 0.624 | 0.000 |
| 0.205 | 0.633 | 0.037 |
| 0.263 | 0.669 | 0.181 |
| 0.411 | 0.685 | 0.238 |
| 0.477 | 0.703 | 0.300 |
| 0.587 | 0.709 | 0.319 |

Table A.1. Lognormal packing fraction η versus standard deviation $\log\sigma_g$ for OS 0.3 as measured by Sohn and Moreland [13], and the computed λ using Eq. (26).

| $\log\sigma_g$ | η | λ |
|---|---|---|
| 0.000 | 0.571 | 0.000 |
| 0.175 | 0.585 | 0.057 |
| 0.267 | 0.602 | 0.122 |
| 0.472 | 0.634 | 0.232 |
| 0.694 | 0.676 | 0.362 |
| 0.915 | 0.693 | 0.410 |
| 1.101 | 0.713 | 0.464 |
| 1.389 | 0.747 | 0.549 |

Table A.2. Lognormal packing fraction η versus standard deviation $\log\sigma_g$ for MR 0.5 as measured by Sohn and Moreland [13], and the computed λ using Eq. (26).

| $\log\sigma_g$ | η | λ |
|---|---|---|
| 0.082 | 0.580 | 0.025 |
| 0.181 | 0.584 | 0.041 |
| 0.406 | 0.617 | 0.164 |
| 0.694 | 0.668 | 0.331 |
| 0.915 | 0.689 | 0.391 |
| 1.098 | 0.703 | 0.432 |
| 1.387 | 0.734 | 0.512 |

Table A.3. Lognormal packing fraction η versus standard deviation $\log\sigma_g$ for MR 1 as measured by Sohn and Moreland [13], and the computed λ using Eq. (26) and an estimated f = 0.574.

| $\log\sigma_g$ | η | λ |
|---|---|---|
| 0.000 | 0.576 | 0.000 |
| 0.082 | 0.580 | 0.017 |
| 0.258 | 0.607 | 0.120 |
| 0.411 | 0.623 | 0.179 |
| 0.694 | 0.670 | 0.332 |
| 0.915 | 0.698 | 0.411 |
| 1.388 | 0.756 | 0.561 |

Table A.4. Lognormal packing fraction η versus standard deviation $\log\sigma_g$ for MR 2 as reported by Sohn and Moreland [13], and the computed λ using Eq. (27).

| $\log\sigma_g$ | η | λ |
|---|---|---|
| 0.00 | 0.59 | 0.000 |
| 0.23 | 0.60 | 0.018 |
| 0.34 | 0.62 | 0.100 |
| 0.46 | 0.63 | 0.131 |
| 0.58 | 0.63 | 0.150 |
| 0.69 | 0.64 | 0.184 |

Table A.5. Lognormal packing fraction η versus standard deviation $\log\sigma_g$ for spheres as reported by Dexter and Tanner [14], and the computed λ using Eq. (26).

| $\log\sigma_g$ | η | λ |
|---|---|---|
| 0.000 | 0.613 | 0.000 |
| 0.048 | 0.621 | 0.035 |
| 0.087 | 0.622 | 0.037 |
| 0.135 | 0.623 | 0.043 |
| 0.192 | 0.629 | 0.064 |
| 0.277 | 0.638 | 0.100 |
| 0.322 | 0.638 | 0.102 |
| 0.390 | 0.646 | 0.131 |
| 0.457 | 0.657 | 0.174 |
| 0.502 | 0.660 | 0.183 |
| 0.703 | 0.690 | 0.289 |

Table A.6. Lognormal packing fraction η versus standard deviation $\log\sigma_g$ for tapped spheres as reported by Suzuki et al. [16], and the computed λ using Eq. (26).

| $\log\sigma_g$ | η | λ |
|---|---|---|
| 0.000 | 0.607 | 0.000 |
| 0.049 | 0.612 | 0.021 |
| 0.090 | 0.613 | 0.024 |
| 0.142 | 0.612 | 0.021 |
| 0.234 | 0.618 | 0.047 |
| 0.283 | 0.629 | 0.087 |
| 0.335 | 0.632 | 0.102 |
| 0.381 | 0.638 | 0.123 |
| 0.444 | 0.645 | 0.150 |
| 0.504 | 0.650 | 0.170 |
| 0.706 | 0.706 | 0.263 |

Table A.7. Lognormal packing fraction η versus standard deviation $\log\sigma_g$ for prodded spheres as reported by Suzuki et al. [16], and the computed λ using Eq. (26).

| $\log\sigma_g$ | η | λ |
|---|---|---|
| 0 | 0.520 | 0 |
| 0.065 | 0.530 | 0.038 |
| 0.132 | 0.532 | 0.047 |
| 0.193 | 0.537 | 0.065 |
| 0.275 | 0.542 | 0.083 |
| 0.335 | 0.550 | 0.113 |
| 0.392 | 0.556 | 0.135 |
| 0.458 | 0.565 | 0.168 |
| 0.507 | 0.573 | 0.193 |
| 0.712 | 0.604 | 0.289 |

Table A.8. Lognormal packing fraction η versus standard deviation $\log\sigma_g$ for tapped crushed glass as reported by Suzuki et al. [16], and the computed λ using Eq. (26).

| $\log\sigma_g$ | η | λ |
|---|---|---|
| 0.000 | 0.505 | 0.000 |
| 0.065 | 0.510 | 0.021 |
| 0.134 | 0.513 | 0.033 |
| 0.199 | 0.517 | 0.048 |
| 0.275 | 0.523 | 0.069 |
| 0.338 | 0.530 | 0.094 |
| 0.395 | 0.538 | 0.125 |
| 0.460 | 0.549 | 0.161 |
| 0.515 | 0.557 | 0.189 |
| 0.714 | 0.590 | 0.291 |

Table A.9. Lognormal packing fraction η versus standard deviation $\log\sigma_g$ for prodded crushed glass as reported by Suzuki et al. [16], and the computed λ using Eq. (27).

| $\log\sigma_g$ | η | λ |
|---|---|---|
| 0.00 | 0.634 | 0.000 |
| 0.01 | 0.634 | 0.003 |
| 0.05 | 0.637 | 0.014 |
| 0.10 | 0.641 | 0.030 |
| 0.15 | 0.643 | 0.040 |
| 0.20 | 0.651 | 0.072 |
| 0.25 | 0.659 | 0.104 |
| 0.30 | 0.668 | 0.141 |
| 0.40 | 0.680 | 0.187 |
| 0.50 | 0.705 | 0.275 |
| 0.60 | 0.715 | 0.310 |
| 0.70 | 0.721 | 0.332 |

Table A.10. Lognormal packing fraction η versus standard deviation $\log\sigma_g$ for spheres as reported by Yang et al. [17], and the computed λ using Eq. (26).

| $\log\sigma_g$ | η | λ |
|---|---|---|
| 0.000 | 0.534 | 0.000 |
| 0.01 | 0.536 | 0.006 |
| 0.05 | 0.540 | 0.024 |
| 0.10 | 0.546 | 0.045 |
| 0.15 | 0.550 | 0.062 |
| 0.20 | 0.558 | 0.090 |
| 0.25 | 0.566 | 0.120 |
| 0.30 | 0.570 | 0.136 |
| 0.40 | 0.586 | 0.189 |
| 0.50 | 0.594 | 0.215 |
| 0.6 | 0.607 | 0.259 |
| 0.7 | 0.619 | 0.293 |

Table A.11. Lognormal packing fraction η versus standard deviation $\log\sigma_g$ for spheres as reported by Yang et al. [17], and the computed λ using Eq. (26).

| $\log\sigma_g$ | η | λ |
|---|---|---|
| 0.000 | 0.628 | 0.000 |
| 0.048 | 0.627 | 0.000 |
| 0.095 | 0.629 | 0.007 |
| 0.152 | 0.634 | 0.028 |
| 0.195 | 0.640 | 0.052 |
| 0.241 | 0.645 | 0.074 |
| 0.295 | 0.655 | 0.111 |
| 0.339 | 0.663 | 0.143 |
| 0.387 | 0.672 | 0.176 |
| 0.438 | 0.680 | 0.208 |
| 0.500 | 0.695 | 0.259 |

Table A.12. Lognormal packing fraction η versus standard deviation $\log\sigma_g$ for spheres as reported by He et al. [18], and the computed λ using Eq. (26).

| $\log\sigma_g$ | η | λ |
|---|---|---|
| 0.00 | 0.644 | 0.000 |
| 0.05 | 0.643 | 0.000 |
| 0.10 | 0.649 | 0.022 |
| 0.15 | 0.652 | 0.034 |
| 0.20 | 0.656 | 0.055 |
| 0.25 | 0.662 | 0.080 |
| 0.30 | 0.671 | 0.116 |
| 0.35 | 0.679 | 0.145 |
| 0.40 | 0.686 | 0.175 |
| 0.425 | 0.695 | 0.208 |
| 0.50 | 0.703 | 0.237 |
| 0.50 | 0.707 | 0.253 |
| 0.525 | 0.712 | 0.269 |
| 0.55 | 0.718 | 0.290 |
| 0.60 | 0.730 | 0.331 |

Table A.13. Lognormal packing fraction η versus standard deviation $\log\sigma_g$ for spheres as reported by Farr and Groot [19], and the computed λ using Eq. (26).

| $\log\sigma_g$ | η | λ |
|---|---|---|
| 0.00 | 0.635 | 0.000 |
| 0.05 | 0.636 | 0.004 |
| 0.10 | 0.639 | 0.015 |
| 0.15 | 0.643 | 0.032 |
| 0.20 | 0.647 | 0.052 |
| 0.25 | 0.654 | 0.079 |
| 0.30 | 0.660 | 0.102 |

Table A.14. Lognormal packing fraction η of spheres at lowest possible packing density ("LT") versus standard deviation $\log\sigma_g$ as reported by Baranau and Tallarek [22], and the computed λ using Eq. (26).

| $\log\sigma_g$ | η | λ |
|---|---|---|
| 0.00 | 0.640 | 0.000 |
| 0.05 | 0.642 | 0.007 |
| 0.10 | 0.644 | 0.015 |
| 0.15 | 0.648 | 0.034 |
| 0.20 | 0.654 | 0.058 |
| 0.25 | 0.660 | 0.082 |
| 0.30 | 0.668 | 0.113 |

Table A.15. Lognormal packing fraction η of spheres at densest possible packing density ("RCP") versus standard deviation $\log\sigma_g$ as reported by Baranau and Tallarek [22], and the computed λ using Eq. (26).

| $\log\sigma_g$ | η | λ |
|---|---|---|
| 0.00 | 0.650 | 0.000 |
| 0.05 | 0.653 | 0.011 |
| 0.10 | 0.655 | 0.023 |
| 0.15 | 0.660 | 0.045 |
| 0.20 | 0.665 | 0.064 |
| 0.25 | 0.672 | 0.092 |
| 0.30 | 0.679 | 0.121 |

Table A.16. Lognormal packing fraction η of spheres at densest glass close packing limit ("GCP") versus standard deviation $\log\sigma_g$ as reported by Baranau and Tallarek [22], and the computed λ using Eq. (27).

| $\log\sigma_g$ | η | λ |
|---|---|---|
| 0.000 | 0.570 | 0.000 |
| 0.100 | 0.577 | 0.026 |
| 0.202 | 0.584 | 0.054 |
| 0.302 | 0.594 | 0.093 |
| 0.402 | 0.604 | 0.131 |
| 0.501 | 0.619 | 0.182 |
| 0.601 | 0.634 | 0.233 |
| 0.701 | 0.653 | 0.294 |
| 0.801 | 0.668 | 0.340 |
| 0.898 | 0.690 | 0.403 |
| 1.000 | 0.706 | 1.000 |

Table A.17. Lognormal packing fraction η of cubes versus standard deviation $\log\sigma_g$ as reported by Malmir et al. [24], and the computed λ using Eq. (26).

| $\log\sigma_g$ | η | λ |
|---|---|---|
| 0.0 | 0.6435 | 0.000 |
| 0.1 | 0.6468 | 0.014 |
| 0.5 | 0.7035 | 0.239 |
| 0.7 | 0.7409 | 0.369 |
| 0.9 | 0.7797 | 0.490 |
| 1.0 | 0.7982 | 0.544 |
| 1.1 | 0.8145 | 0.589 |
| 1.25 | 0.8332 | 0.639 |
| 1.3 | 0.8435 | 0.665 |

Table A.18. Lognormal packing fraction η versus standard deviation $\log\sigma_g$ for spheres as reported by Desmond [25, 42], and the computed λ using Eq. (26).

| $\log\sigma_g$ | η | λ |
|---|---|---|
| 0.00 | 0.6435 | 0.000 |
| 0.05 | 0.6447 | 0.005 |
| 0.1 | 0.6473 | 0.016 |
| 0.15 | 0.6515 | 0.034 |
| 0.2 | 0.6571 | 0.058 |
| 0.25 | 0.664 | 0.086 |
| 0.3 | 0.6717 | 0.118 |
| 0.35 | 0.6801 | 0.151 |
| 0.4 | 0.6888 | 0.184 |
| 0.45 | 0.6981 | 0.219 |
| 0.5 | 0.7076 | 0.254 |
| 0.55 | 0.7172 | 0.288 |
| 0.6 | 0.727 | 0.322 |
| 0.65 | 0.7369 | 0.355 |
| 0.7 | 0.7466 | 0.387 |
| 0.75 | 0.7562 | 0.418 |
| 0.8 | 0.7656 | 0.447 |
| 0.85 | 0.775 | 0.476 |
| 0.9 | 0.7841 | 0.503 |
| 0.95 | 0.7932 | 0.529 |
| 1.0 | 0.8018 | 0.554 |
| 1.05 | 0.8102 | 0.577 |
| 1.1 | 0.8181 | 0.599 |
| 1.15 | 0.8262 | 0.620 |
| 1.2 | 0.8339 | 0.640 |
| 1.25 | 0.8412 | 0.659 |
| 1.3 | 0.8484 | 0.677 |
| 1.4 | 0.8617 | 0.710 |
| 1.5 | 0.8741 | 0.740 |
| 1.6 | 0.8854 | 0.766 |
| 1.7 | 0.8964 | 0.791 |
| 1.8 | 0.9062 | 0.813 |
| 1.9 | 0.9147 | 0.832 |
| 2.0 | 0.9235 | 0.850 |
| 2.1 | 0.9301 | 0.864 |
| 2.2 | 0.9374 | 0.879 |
| 2.3 | 0.9431 | 0.891 |
| 2.4 | 0.9489 | 0.903 |
| 2.5 | 0.954 | 0.913 |
| 2.6 | 0.9584 | 0.922 |
| 2.7 | 0.9621 | 0.929 |
| 2.8 | 0.9669 | 0.938 |
| 2.9 | 0.9699 | 0.944 |
| 3.0 | 0.9728 | 0.950 |
| 3.1 | 0.9762 | 0.956 |
| 3.2 | 0.9786 | 0.961 |
| 3.3 | 0.9809 | 0.965 |
| 3.4 | 0.9817 | 0.966 |
| 3.5 | 0.9841 | 0.971 |

Table A.19. Lognormal packing fraction η versus standard deviation $\log\sigma_g$ for spheres as reported by Desmond [42], using the algorithm of [19], and the computed λ using Eq. (26).

| $\log\sigma_g$ | η | λ |
|---|---|---|
| 0 | 0.8425 | 0.000 |
| 0.2 | 0.8482 | 0.043 |
| 0.3 | 0.8529 | 0.077 |
| 0.4 | 0.8594 | 0.125 |
| 0.5 | 0.867 | 0.179 |
| 0.6 | 0.8752 | 0.237 |
| 0.7 | 0.8842 | 0.299 |
| 0.8 | 0.893 | 0.359 |
| 0.9 | 0.9029 | 0.425 |
| 1 | 0.9115 | 0.481 |
| 1.1 | 0.9194 | 0.531 |
| 1.2 | 0.9271 | 0.579 |
| 1.3 | 0.9335 | 0.619 |
| 1.4 | 0.94 | 0.659 |
| 1.5 | 0.9457 | 0.693 |
| 1.6 | 0.9492 | 0.714 |

Table A.20. Lognormal packing fraction η versus standard deviation $\log\sigma_g$ for disks in a plane as reported

by Desmond [25, 42], and the computed λ using Eq. (26). The monosized packing fraction is taken from [32].

| $\log\sigma_g$ | η | λ |
|---|---|---|
| 0.0 | 0.8425 | 0.000 |
| 0.1 | 0.8460 | 0.026 |
| 0.2 | 0.8487 | 0.046 |
| 0.3 | 0.8537 | 0.083 |
| 0.4 | 0.8598 | 0.128 |
| 0.5 | 0.8670 | 0.179 |
| 0.6 | 0.8749 | 0.235 |
| 0.8 | 0.8915 | 0.349 |
| 0.9 | 0.9001 | 0.406 |
| 1.0 | 0.9081 | 0.459 |
| 1.2 | 0.9219 | 0.547 |
| 1.4 | 0.9337 | 0.620 |
| 1.5 | 0.9346 | 0.626 |

Table A.21. Lognormal packing fraction η versus standard deviation $\log\sigma_g$ for disks in a plane as reported by Weeks [41], and the computed λ using Eq. (26). The monosized packing fraction is taken from [32].

**DATA AVAILABILITY**

The data that support the findings of this article are not publicly available. The data are available from the author upon reasonable request.

---

[1] S. Torquato and F.H. Stillinger, Jammed hard-sphere packings: from Kepler to Bernal and beyond, Rev. Mod. Phys. 82, 2633 (2010).
[2] P. Charbonneau, A. Ikeda, A., G. Parisi, G. and F. Zamponi, Glass transition and random close packing above three dimensions, Phys. Rev. Lett. 107, 185702 (2011).
[3] G. Parisi and F. Zamponi, Mean-field theory of hard sphere glasses and jamming, Rev. Mod. Phys. 82, 789 (2010).
[4] F.M. Archibald, Particle packing, MSc thesis, University of Toronto (1926).
[5] C.C. Furnas, The relation between specific volume, voids, and size composition in systems of broken solids of mixed sizes Department of Commerce, Bureau of Mines, Report of Investigation Serial No. 2894 (1928).
[6] A.E.R. Westman and H.R. Hugill, The packing of particles, J. Am. Ceram. Soc. 13, 767 (1930).
[7] A.E.R. Westman, The packing of particles: empirical equations for intermediate diameter ratios, J. Am. Ceram. Soc. 19, 127 (1936).
[8] M.A. Caquot, Le rôle des materiaux inertes dans le béton, Mémoires de la Société des Ingénieurs Civils de France, 562 (in French) (1937).
[9] P.C. Mangelsdorf and E.L. Washington, Nature, Lond. 187, 930 (1960).
[10] R.K. McGeary, J. Am. Ceram. Soc. 44, 513 (1961).
[11] R. Jeschar, W. Pötke, V. Petersen and K. Polthier, Proceedings Symposium on Blast Furnace Aerodynamics, Wollongong, Australia, 25-27 September 1975, Ed. N. Standish, Illawara Branch of the Australasian Institute of Mining and Metallurgy (Aus. I.M.M.) (1975), pp. 136-147.
[12] W. Luck, M. Klier and H. Wesslau, Ueber Bragg-Reflexe mit sichtbarem Licht an monodispersen Kunststofflatices I, Berichte der Bunsengesellschaft 67, 75 (in German) (1963).
[13] H.Y. Sohn and C. Moreland, The effect of particle size distribution on packing density, Canadian J. Chem. Eng. 46, 162 (1968).
[14] A.R. Dexter and D.W. Tanner, Packing density of ternary mixtures of spheres, Nature 238, 31 (1972).
[15] P.N. Pusey and W. van Megen, Phase behaviour of concentrated suspensions of nearly hard colloidal spheres, Nature 320, 340 (1986).
[16] M. Suzuki, T. Oshima, H. Ichiba and I. Hasegawa, Void fraction of multi-component randomly packed beds with size distributions, KONA Powder and Particle J. 4, 4 (1986).
[17] A. Yang, C.Y. Miller and L.D. Turcoliver, Simulation of correlated and uncorrelated packing of random size spheres, Phys. Rev. E 53, 1516 (1996).
[18] D. He, N.N. Ekere and L. Cai, Computer simulation of random packing of unequal particles, Phys. Rev. E 60, 7098 (1999).
[19] R.S. Farr and R.D. Groot, Close packing density of polydisperse hard spheres, J. Chem. Phys. 131, 244104 (2009).
[20] M. Clusel, E.I., Corwin, A.O.N. Siemens and J. Brujić, A ‘granocentric’ model for random packing of jammed emulsions, Nature 460, 611 (2009).
[21] K.W. Desmond and E.R. Weeks, Influence of particle size distribution on random close packings of spheres, Phys. Rev. E. 90, 022204 (2014).
[22] V. Baranau and U. Tallarek, Random-close packing limits for monodisperse and polydisperse hard spheres, Soft Matter 10, 3826 (2014).
[23] E. Zaccarelli, S.M. Liddle and W.C.K. Poon, On polydispersity and the hard sphere glass transition, Soft Matter 11, 324 (2014).
[24] M. Malmir, M. Sahimi and M. Reza Rahimi Tabar, Statistical characterization of microstructure of packings of polydisperse hard cubes, Phys. Rev. E 95, 052902 (2017).
[25] K.W. Desmond, Random close packing at extreme size ratios with an Adam-based inflation protocol, arXiv:2608.12235v1 (2026).
[26] I. Biazzo, F. Caltagirone, G. Parisi and F. Zamponi, Theory of Amorphous Packings of Binary Mixtures of Hard Spheres, Phys. Rev. Lett. 102, 195701 (2009).
[27] H. Ikeda, K. Miyazaki, H. Yoshino, A. Ikeda, Multiple glass transitions and higher-order replica symmetry breaking of binary mixtures, Phys. Rev. E 103, 022613 (2021).
[28] C.L.C. Laudicina, I. Pihlajamaa and L.M.C. Janssen, Competing relaxation channels in continuously

polydisperse fluids: A mode-coupling study, Phys. Rev. Res. 5, 033121 (2023).
[29] H. Kim and A. Ikeda, Replica theory for the dynamic glass transition of hard spheres with continuous polydispersity, Phys. Rev. E 111, 035401 (2025).
[30] R. Schlothauer and T. Franosch, Mode-coupling theory of the glass transition for polydisperse systems, Phys. Rev. E. 104, 025412 (2026).
[31] H.J.H. Brouwers, Random packing fraction of binary similar particles: Onsager's excluded volume model revisited, Physics – Uspekhi 67, 510 (2023).
[32] H.J.H. Brouwers, Random packing fraction of binary hyperspheres with small or large size difference: A geometric approach, Phys. Rev. E 112, 025411 (2025).
[33] L. Onsager, The effects of shape on the interaction of colloidal particles, Ann. N.Y. Acad. Sci. 51, 627 (1949).
[34] H.J.H. Brouwers, Excluded volume geometry and packing fraction in binary convex hyperparticle mixtures, Phys. Rev. E 113, 025417 (2026).
[35] J. Aitchison and J.A.C. Brown, The lognormal distribution (Cambridge University Press, Cambridge, 1957).
[36] E. Limpert, W.A. Stahel and M. Abbt, Log-normal distributions across the sciences: keys and clues, BioScience 53, 341 (2001).
[37] A.N. Kolmogoroff, Über das logarithmisch normale Verteilungsgesetz der Dimensionen der Teilchen bei Zerstückelung. Comptes rendu (Doklady) de l'Académie des Sciences de l'USSR 31, 99 (in German) (1941).
[38] J. Söderlund, L.B. Kiss, G. A. Niklasson, and C. G. Granqvist, Lognormal size distributions in particle growth processes without coagulation Phys. Rev. Lett. 80, 2386 (1998).
[39] W. Feller, An introduction to probability theory and its applications, Vols. I (3rd ed.) and II (2nd ed.) (John Wiley, New York, 1970).
[40] K.W. Desmond, Private communications (2026).
[41] E. Weeks, Private communications (2026).
[42] K.W. Desmond and E.R. Weeks, Random close packing of disks and spheres in confined geometries, Phys. Rev. E. 80, 051305 (2009).